\documentclass[twocolumn]{aastex631}

\usepackage[english]{babel}
\usepackage{amsmath}
\usepackage{amssymb}
\usepackage{graphicx}
\usepackage{subfigure}
\usepackage[normalem]{ulem}
\usepackage{enumerate}
\usepackage{booktabs}

\usepackage{verbatim}

\usepackage{color}
\usepackage{multirow}
\usepackage{mathtools}
\usepackage{epstopdf}
\usepackage{tabularx}

\usepackage{url}
\usepackage{xcolor}

\def\apjl{ApJL}
\def\apjs{ApJS}
\def\aap{A\&A}

\def\be{\begin{equation}} 
\def\ee{\end{equation}} 
\def\ba{\begin{eqnarray}} 
\def\ea{\end{eqnarray}}

\def\mpc{\,{\rm {Mpc}}}

\def\msun{{\Msun}}

\def\gsim{\lower.5ex\hbox{\gtsima}} 
\def\lsim{\lower.5ex\hbox{\ltsima}} \def\gtsima{$\; \buildrel > \over 
\sim \;$} \def\ltsima{$\; \buildrel < \over \sim \;$} \def\prosima{$\; 
\buildrel \propto \over \sim \;$} \def\gsim{\lower.5ex\hbox{\gtsima}} 
\def\lsim{\lower.5ex\hbox{\ltsima}} 
\def\simgt{\lower.5ex\hbox{\gtsima}} 
\def\simlt{\lower.5ex\hbox{\ltsima}} 
\def\simpr{\lower.5ex\hbox{\prosima}}   
  
 \def\gtsima{$\; \buildrel > \over \sim \;$} 
\def\ltsima{$\; \buildrel < \over \sim \;$} 
\def\gsim{\lower.5ex\hbox{\gtsima}} 
\def\lsim{\lower.5ex\hbox{\ltsima}} 
\def\simgt{\lower.5ex\hbox{\gtsima}} 
\def\simlt{\lower.5ex\hbox{\ltsima}} 
\def\simpr{\lower.5ex\hbox{\prosima}}

\def\msun{\,{\rm \Msun}}

\def\E3{{\cal E}_{\rm g}^{III}}

\def\Msun{\rm M_\odot}

\def\r12{r_{1/2}} 
\def\x12{x_{1/2}} 
\def\v12{v_{1/2}}

\newcommand\code[1]{\textsc{\MakeLowercase{#1}}}

\def\nh2{n_{\rm H2}}
\def\fh2{f_{\rm H2}}

\def\angstrom{\textrm{A\kern -1.3ex\raisebox{0.6ex}{$^\circ$}}}

\def\msun{{\rm M}_{\odot}}
\def\zsun{{\rm Z}_{\odot}}

\def\msunyr{\msun\,{\rm yr}^{-1}}

\def\kpc{{\rm kpc}}

\makeatletter
\def\@hex@@Hex#1%
 {\if a#1A\else \if b#1B\else \if c#1C\else \if d#1D\else
  \if e#1E\else \if f#1F\else #1\fi\fi\fi\fi\fi\fi \@hex@Hex}
\makeatother

\definecolor{apcolor}{HTML}{b3003b}
\definecolor{cbcolor}{HTML}{ff0f00}
\definecolor{afcolor}{HTML}{b3443c}
\definecolor{vgcolor}{HTML}{8F00FF}
\definecolor{tbdcolor}{HTML}{E8A95E}
\definecolor{stefcolor}{HTML}{0047ab}

\shorttitle{Quiescent galaxies with JWST}
\shortauthors{Gelli et al.}

\begin{document}

\title{Quiescent low-mass galaxies observed by JWST in the Epoch of Reionization}

\correspondingauthor{Viola Gelli}
\email{viola.gelli@unifi.it}

\author[0000-0001-5487-0392]{Viola Gelli}
\author[0000-0001-7298-2478]{Stefania Salvadori}
\affiliation{Dipartimento di Fisica e Astronomia, Universit\'{a} degli Studi di Firenze, via G. Sansone 1, 50019, Sesto Fiorentino, Italy}
\affiliation{INAF/Osservatorio Astrofisico di Arcetri, Largo E. Fermi 5, I-50125, Firenze, Italy}
\author[0000-0002-9400-7312]{Andrea Ferrara}
\author[0000-0002-7129-5761]{Andrea Pallottini}
\author[0000-0002-6719-380X]{Stefano Carniani}
\affiliation{Scuola Normale Superiore, Piazza dei Cavalieri 7, I-56126 Pisa, Italy}

\begin{abstract}
The surprising JWST discovery of a quiescent, low-mass ($M_\star=10^{8.7} \msun$) galaxy at redshift $z=7.3$ (JADES-GS-z7-01-QU) represents a unique opportunity to study the imprint of feedback processes on early galaxy evolution. We build a sample of 130 low-mass ($M_\star\lesssim 10^{9.5} \msun$) galaxies from the \code{serra} cosmological zoom-in simulations, which show a feedback-regulated, bursty star formation history (SFH). The fraction of time spent in an active phase increases with the stellar mass from $f_{duty}\approx 0.6$ at $M_\star\approx 10^{7.5}\msun$ to $\approx 0.99$ at $M_\star\geq 10^{9}\msun$, and it is in agreement with the value $f_{duty}\approx 0.75$ estimated for JADES-GS-z7-01-QU. On average, 30\% of the galaxies are quiescent in the range $6 < z < 8.4$; they become the dominant population at $M_\star\lesssim 10^{8.3} \msun$. However, none of these quiescent systems matches the Spectral Energy Distribution of JADES-GS-z7-01-QU, unless their SFH is artificially truncated a few Myr after the main star formation peak. As supernova feedback can only act on a longer timescale ($\gtrsim 30$ Myr), this implies that the observed abrupt quenching must be caused by a faster physical mechanism, such as radiation-driven winds.
\end{abstract}


\section{Introduction}
The growth of galaxies is regulated by the conversion of cold gas into stars and is affected by different feedback processes \citep{CiardiFerrara05}, which can lead to the temporary or definitive quenching of the star formation activity. Internal feedback mechanisms include radiative feedback, which can destroy H$_2$ molecules preventing gas cooling \citep[e.g.][]{Johnson2007,Krumholz09}, and mechanical feedback from massive stars/supernovae (SN) or active galactic nuclei (AGN), which can partially or completely remove the gas reservoir \citep[e.g.][]{MacLowFerrara1999,Croton2016, Carnall23}. 

The evolution of low-mass galaxies, which we define here as those with stellar mass $M_\star\leq 10^{9.5}\msun$, is dramatically affected, regulated, and -- in some cases -- even halted by these physical mechanisms because of their shallow potential well \citep[e.g.][]{FerraraTolstoy01,Salvadori08,Wise12,collins2022}. Furthermore, these systems can also be impacted by more global feedback mechanisms, such as reionization, preventing the accretion of fresh gas into the less massive systems \citep[e.g.][]{gnedin2000,Dijkstra2004,sobacchi2013,Pereira-Wilson2023}, and environmental effects, such as ram-pressure stripping, which can remove the interstellar medium from satellite galaxies \citep[e.g.][]{Meyer2006,Emerick2016,Boselli2022}. Thus, low-mass galaxies are the ideal laboratory to investigate feedback processes.

\begin{figure*}
\centering
\includegraphics[width=\textwidth]{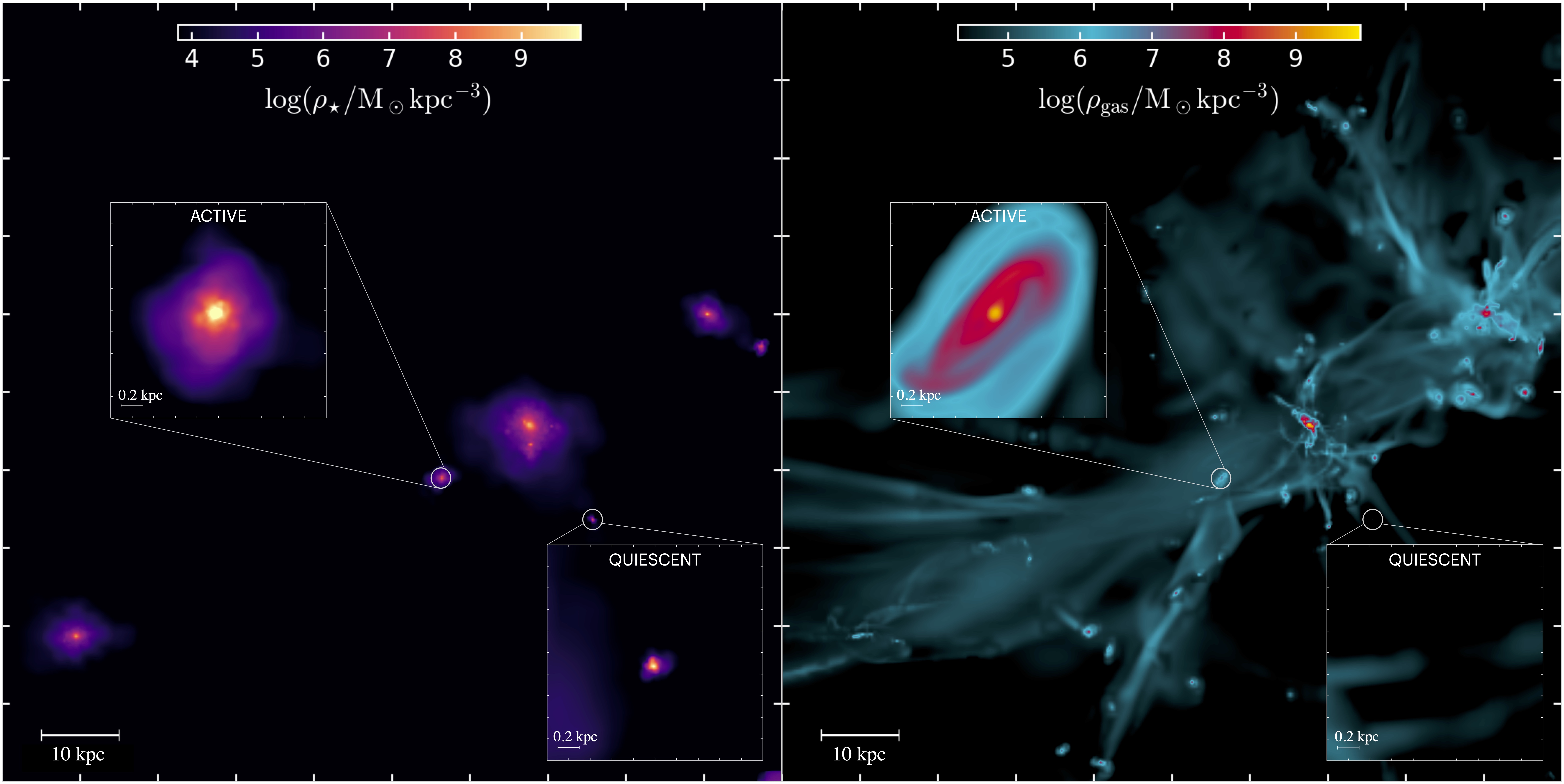}
\caption{Example of stellar ($\rho_\star$, left) and gas ($\rho_{\rm gas}$, right) density maps of a system at $z=6$ within the \code{SERRA} simulations.
The field of view of $\rm 100~kpc \times 100~kpc$ contains six galaxies. The insets show two zoomed regions, each of size $\rm 2~kpc \times 2~kpc$, centred on a typical quiescent  ($\log M_\star / \msun = 7.96 $) and active ($\log M_\star / \msun = 8.97 $) galaxy, as indicated.
\label{fig:maps}
}
\end{figure*}

In the early Universe, low-mass galaxies are very common. Thanks to JWST we can now study them and investigate the role of the various feedback processes in their evolution. \cite{Looser23} recently reported the discovery of a quiescent low-mass galaxy at redshift $z=7.3$ (see also \citealt{Strait23} for another very low-mass quiescent system at $z=5.2$) from the JADES program, JADES-GS-z7-01-QU.
Following this work, we define as {\it quiescent} those galaxies with no (or negligible) star formation activity at the epoch of observation\footnote{Note that in the literature quiescent galaxies are sometimes differently defined as those below the main sequence \citep{houston2023}, or quenched for $>500$~Myr \citep{Perez23}.}.
According to \cite{Looser23}, the spectral energy distribution (SED) of JADES-GS-z7-01-QU is consistent with a metal-poor $M_\star \approx 5\times 10^8\msun$ stellar population formed in a short and intense burst of star formation followed by rapid quenching (star formation rate $\rm SFR < 10^{-2.5}\msunyr$), about 10–20 Myr before the epoch of observation.
Given the rapidity of the transition from a star-forming to a quiescent state for JADES-GS-z7-01-QU,
its quenching is expected to have been driven by a fast process. A reasonable explanation may therefore be associated with the injection of mechanical energy through outflows by either star formation or AGN.

In this Letter, we aim at interpreting these new findings and address the following questions: what is the duty cycle of high-redshift low-mass galaxies? When and why low-mass galaxies become quiescent? What are the key physical mechanisms causing their quenching?

\section{Simulated low-mass galaxies}\label{sec:methods}
To answer the above questions, we use the \code{SERRA} \citep{Pallottini2022} suite of high-resolution cosmological zoom-in simulations that follow the evolution of typical Lyman break galaxies during the Epoch of Reionization. 
The simulations evolve from $z=100$, where initial conditions are generated with \code{Music} \citep{HahnAbel11}, in a cosmological volume of $(20\mpc /h)^3$ by assuming a \citet{Planck14} cosmology\footnote{Throughout the paper we assume a $\Lambda$CDM model with vacuum, matter, and baryon densities in units of the critical density $\Omega_{\Lambda}= 0.692$, $\Omega_{m}= 0.308$, $\Omega_{b}= 0.0481$, Hubble constant $\rm H_0=67.8\, {\rm km}\,{\rm s}^{-1}\,{\rm Mpc}^{-1}$, spectral index $n=0.967$, and $\sigma_{8}=0.826$.}. 
A customised version of the Adaptive Mesh Refinement code \code{RAMSES} \citep{Teyssier02} is used for the evolution of dark matter (DM), stars and gas, reaching a baryon mass resolution of $1.2 \times 10^4 \Msun$ and spatial resolution of $\simeq 20~\,\rm pc$ in the zoom-in regions, i.e. about the mass and size of molecular clouds.

On-the-fly radiative transfer is included through \code{RAMSES-RT} \citep{Rosdahl2013}, and a non-equilibrium chemical network generated with \code{KROME} \citep{krome} is used for regulating the interaction of the gas with photons \citep{Pallottini17}.
Stars form according to a Schmidt–Kennicutt relation \citep{Kennicutt98} depending on the molecular-hydrogen gas density, and assuming a \cite{Kroupa01} initial mass function for the stellar particles. Stellar feedback modelling includes SNe explosions, winds and radiation pressure \citep{Pallottini16}. The energy inputs and chemical yields, depending on the stellar age and metallicity, are computed through \code{STARBURST99} \citep{starburst99} using \code{PADOVA} stellar tracks \citep{Bertelli94}, covering a metallicity range of $Z_\star / \rm Z_\odot = 0.02-1.0$. Since the resolution does not allow to follow the evolution of the first stars and mini-haloes, their effect on the ISM is reproduced by setting the initial gas metallicity to a floor value of $Z_{\rm floor}=10^{-3} Z_\odot$ \citep{Wise12,Pallottini14}.

The emission of the galaxies is modelled using \code{STARBURST99} for the stellar and nebular continuum \citep[see also][]{Gelli21}. We use \code{CLOUDY} \citep{Ferland17} to compute nebular line emission (considering the main ones typically contributing to the rest-frame UV-optical spectrum, i.e. $\rm H\alpha$, $\rm H\beta$, $\rm H\gamma$, [OII]$\lambda\lambda$3726,3729, [OIII]$\lambda\lambda$4959,5007 and CIII]1909), accounting for the ISM density, metallicity, internal structure, and radiation field \citep[e.g.][]{vallini:2018,Pallottini19,Kohandel19}.
Finally, we take into account the presence of dust which attenuates the intrinsic galaxy spectrum \citep[i.e.][]{Gelli21}, adopting a dust-to-metal ratio $f_d = 0.08$ and assuming a MW-like dust composition and grain size distribution\footnote{This choice is driven by ALMA measurements of high-$z$ galaxies \citep{bouwens:2022}, that support a MW-like dust \citep{Ferrara22} and low dust-to-metal ratios \citep[][]{Behrens18, Laporte17a}. However, the use use different extinction curves (e.g. SMC, LMC) does not lead to significant changes in the emission of low-mass galaxies.} \citep{Weingartner01}.

\begin{figure}[t!]
\centering
\includegraphics[width=0.46\textwidth]{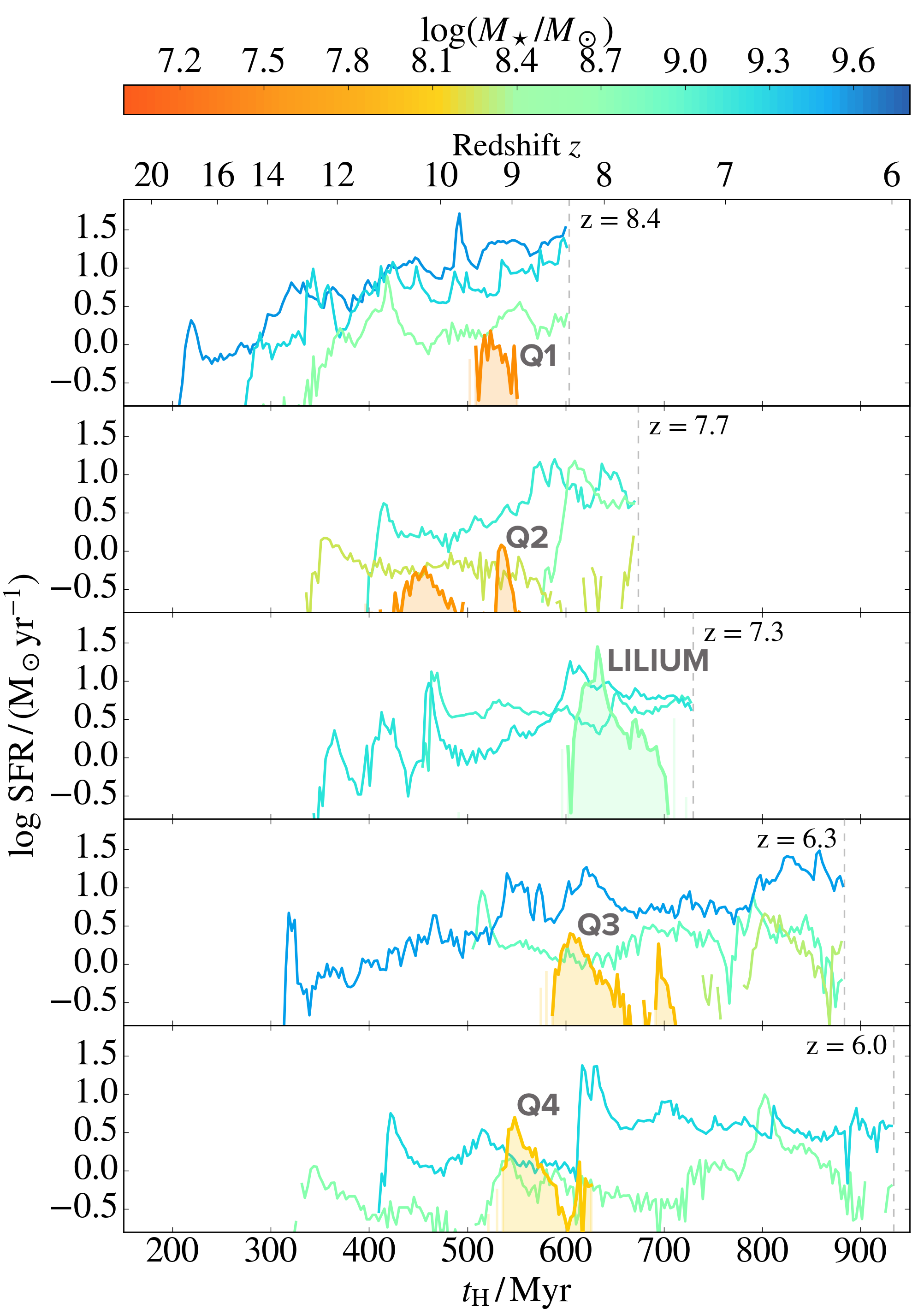}
\caption{Star formation rate (SFR) as a function of the age of the Universe ($t_H$) for some \code{SERRA} galaxies simulated up to different redshifts, color coded with the final stellar mass. Filled curves and names denote galaxies that are quiescent at the end of the simulations (30\% of the sample).
\label{fig:histories}
}
\end{figure}

We analyse multiple snapshots of different \code{SERRA} runs in the range $6<z<8.4$. We follow the stellar-density method implemented in \cite{Gelli20} to select low-mass galaxies with $M_\star \lesssim 10^{9.5} \Msun$. The final sample has 130 galaxies.

As an example, in Fig.~\ref{fig:maps} we show stellar and gas density maps from one of the simulations at $z=6$. The displayed volume ($10^6~\kpc^3$) contains 6 galaxies. In the two insets we zoom on typical low-mass systems: an actively star-forming ($\log M_\star / \msun = 8.97$), and a {quiescent} system with no ongoing star formation ($\log M_\star / \msun = 7.96$). The former shows an extended ($\sim 500 \rm~pc$ in radius) stellar distribution, and a gas proto-disk structure, typical of more massive galaxies. The quiescent galaxy is instead smaller, both in terms of stellar mass and size, with all of its stars concentrated in the inner $\lesssim 200 \rm~pc$, and almost completely devoid of gas. 

\begin{figure}[t!]
\centering
\includegraphics[width=0.46\textwidth]{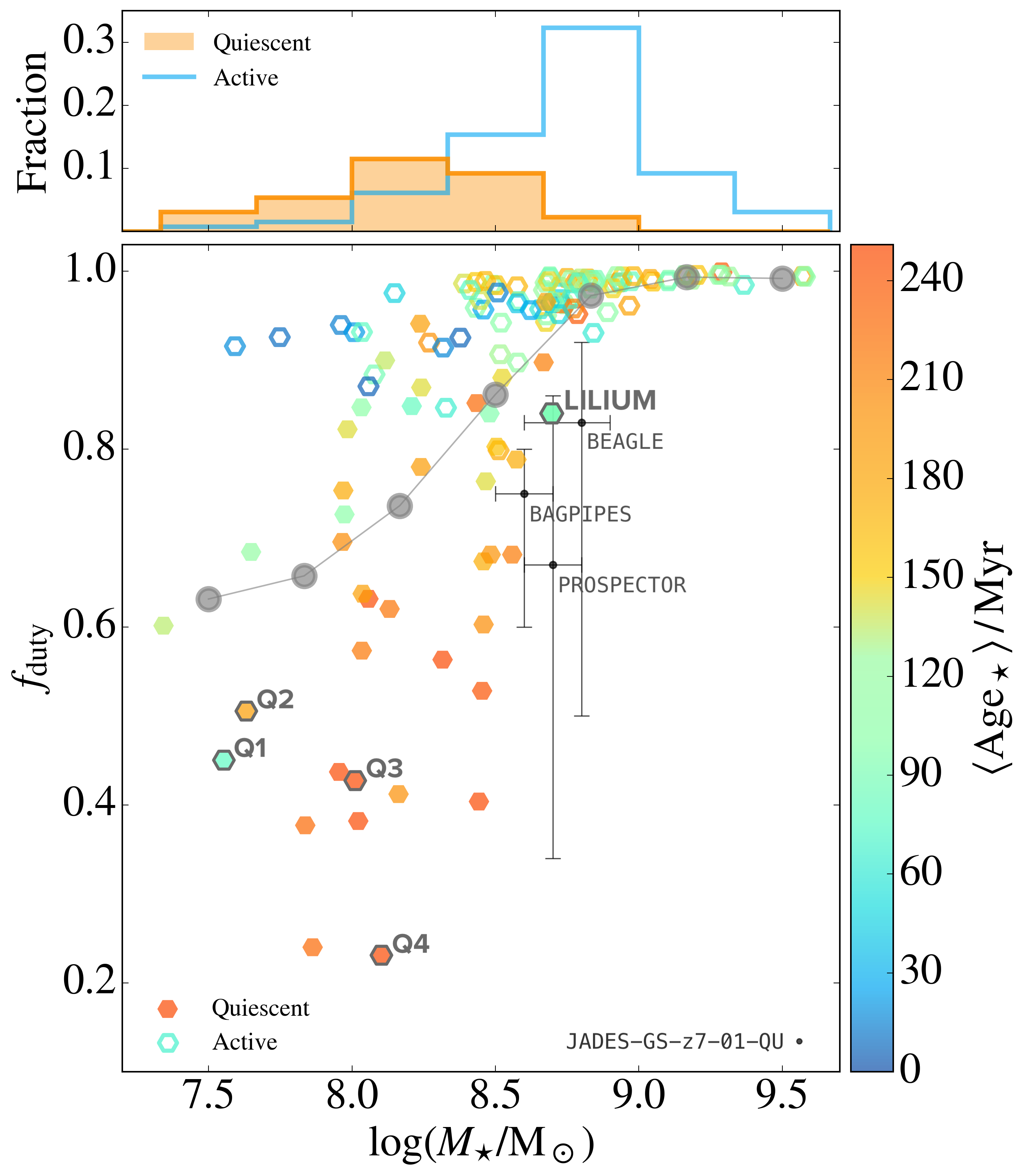}
\caption{Duty cycle $f_{\rm duty}$ as a function of the stellar mass of quiescent (filled) and active (hollow) galaxies, coloured with the average age of the stellar population.
The average trend of $f_{\rm duty}$ with the mass is shown through the grey circles.
The three points with errorbars are the values inferred for JADES-GS-z7-01-QU using different spectral fitting codes (\code{BAGPIPES} by \citealt{carnall_bagpipes}, \code{PROSPECTOR} by \citealt{prospector21}, \code{BEAGLE} by \citealt{chevallard_beagle}) derived from Table~1 of \cite{Looser23}.
The histograms in the upper panels show the stellar mass distribution of active and quiescent low-mass galaxies.
\label{fig:dutycycle}
}
\end{figure}

\section{Nature of quiescent systems}

In Fig.~\ref{fig:histories} we show the star formation histories (SFHs) in terms of age of the Universe $t_{\rm H}$ for some \code{SERRA} low-mass galaxies up to different observation redshifts. The SFHs have an intermittent nature made of bursts followed by star formation drops or even halts. Such behaviour results from stellar feedback, efficiently decreasing or even suppressing star formation in low-mass galaxies. Note that all galaxies with final mass $M_\star \lesssim 10^{9} \Msun$ have experienced quiescent SF phases, or are still quenched at the final snapshot, which is identified as the redshift of observation. On average, these quenched galaxies represent the 30\% of the population. 

Interestingly, after the SF peak, these low-mass galaxies feature a smooth SFR decrease lasting some tens of Myr until the final quenching. This trend is typical of stellar feedback powered by SNe, as explosions occur with a delay time that increases with a decreasing SN progenitor mass \citep[see][]{Gelli20}.
This behaviour is encountered in all low-mass systems, i.e. in both field and satellite galaxies, implying that environmental processes do not drive their evolution.

In the middle panel of Fig.~\ref{fig:histories} we pinpoint the galaxy {\it Lilium}, which is quiescent at $z=7.3$ and has the same stellar mass of JADES-GS-z7-01-QU ($10^{8.7}\msun$).

Fig.~\ref{fig:dutycycle} shows the galaxy SFR duty cycle, i.e. the ratio between the actively star forming (SFR$>0$) time interval, $\Delta t_{\rm on}$, and the time elapsed between the first star formation event, $t_{\rm form}$, and the observation redshift at $t_{\rm obs}$: 
\begin{equation}
f_{\rm duty} = \frac{\Delta t_{\rm on}}{t_{\rm obs}- t_{\rm form}}\,.
\label{eq:dc}
\end{equation}
Galaxies are color-coded with the mass-averaged age of their stellar populations; quiescent systems (Q1-Q4) identified in Fig.~\ref{fig:histories} are specifically indicated. 

Galaxies with $M_\star < 10^{9}\msun$ can be either quiescent or active, and the fraction of quiescent systems increases with decreasing stellar mass, becoming the dominating population for $M_\star < 10^{8.3}\msun$ (see histograms in the upper panel of Fig.~\ref{fig:dutycycle}). In this mass range quiescent galaxies (a) are older (average ages $>100$~Myr), (b) have lower and more scattered duty cycles ($f_{duty}\approx 0.2-0.9$) than active galaxies ($f_{duty}\approx 0.8-0.99$).
For $M_\star > 10^{9}\msun$, instead, {\it all} galaxies are active, and they have been forming stars for $\geq 95 \%$ of their lifetime. 

The average duty cycle of quiescent and active galaxies clearly reflects these trends with stellar mass, slowly increasing from $f_{duty}\approx 0.6$ for $M_\star\approx 10^{7.5}\msun$ to $\approx 0.99$ for $M_\star\geq 10^{9}\msun$.
Noticeably, the quiescent galaxy {\it Lilium} ($\log M_\star / \msun = 8.7$) has a duty cycle $f_{\rm duty}\approx 0.84$ that is consistent with the values obtained for JADES-GS-z7-01-QU using different stellar population synthesis tools by \citealt{Looser23}. For instance, when considering the values derived with \code{BAGPIPES} in Table 1 therein\footnote{Note that in \cite{Looser23} $t_{\rm form}$ is defined as lookback time from the observation, while in this paper it is defined in terms of age of the Universe.}, the time from the first star formation event is $t_{\rm obs}- t_{\rm form} \approx 40$~Myr, and the time of active star formation is $\Delta t_{\rm on} \approx 30$~Myr. This implies a duty cycle (see Equation~\ref{eq:dc}) for JADES-GS-z7-01-QU of $f_{\rm duty}\approx 0.75$.

\begin{figure*}
%\centering
\includegraphics[width=0.98\textwidth]{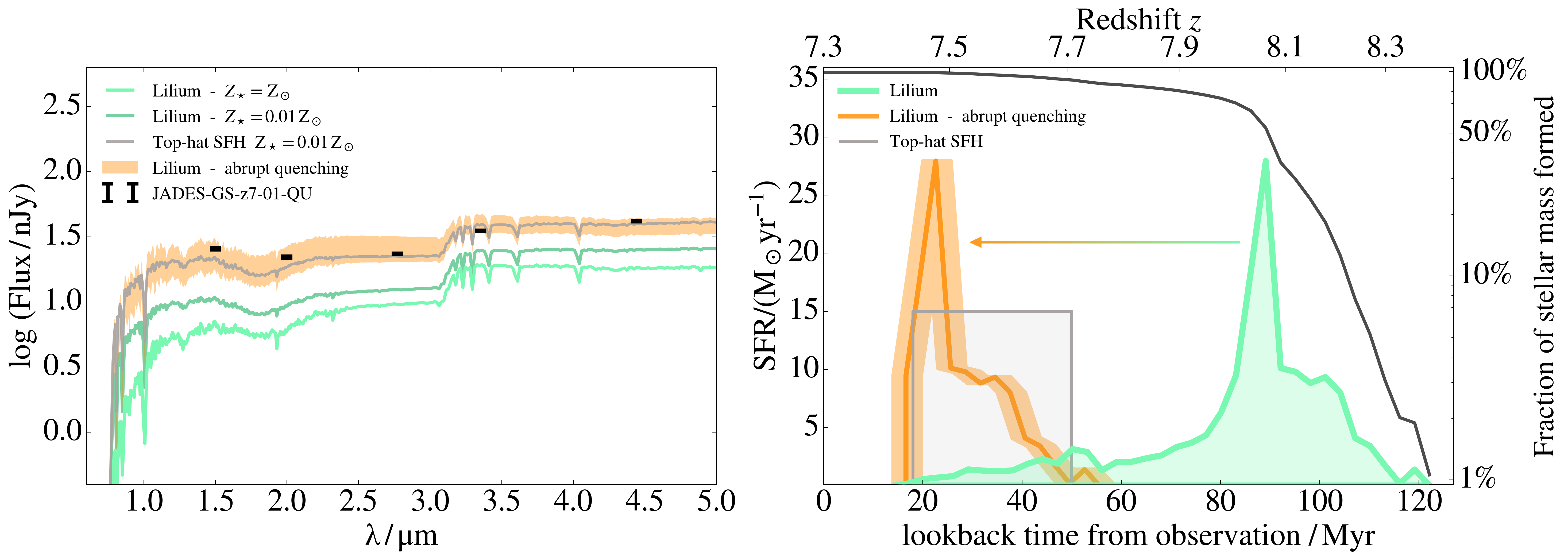}
\caption{Spectral energy distributions of simulated galaxies at $z=7.3$ compared with JADES-GS-z7-01-QU (\textit{left panel}), and corresponding star formation histories (\textit{right}). Shown in the two figures are the SEDs and SFHs for: (a) {\it Lilium} with two stellar metallicities as indicated (green), (b) an idealised top-hat SFH model with constant SFR (grey), and (c) {\it Lilium} with a modified SFH featuring an abrupt quenching $\approx 5$~Myr ater the main peak of star formation and observed after $\Delta t_{\rm quench}$ varying between $\rm 15-20~Myr$ (orange). The black curve in the right panel shows the fraction of stellar mass formed for {\it Lilium}.
\label{fig:sed_SFR_Z}
}
\end{figure*}

\section{Comparison with JWST observations}\label{sec:obs_comp}
{\it Lilium} was selected among \code{SERRA} quiescent galaxies due to its similarity with JADES-GS-z7-01-QU in terms of stellar mass, duty cyle (Fig. \ref{fig:dutycycle}), and the time elapsed between SF quenching and observation redshift ($\Delta t_{\rm quench} \sim 15$~Myr for {\it Lilium} vs. $\Delta t_{\rm quench}\sim 10 - 40$~Myr for JADES-GS-z7-01-QU). However, the stellar populations of {\it Lilium} are more metal-rich ($Z_\star \approx \zsun$) than deduced by \citet{Looser23} for JADES-GS-z7-01-QU, $Z_\star \approx 0.01 \,\zsun$. In spite of these similarities, the SED we predict for {\it Lilium} (Fig.~\ref{fig:sed_SFR_Z}, left panel) is too faint to match the one observed in JADES-GS-z7-01-QU. What is the origin of such discrepancy?

We first check the role of stellar metallicity by imposing that all the stars in {\it Lilium} have $Z_\star \approx 0.01 \, \zsun$. However, the resulting SED only increases by $\simlt 0.2$ dex (see Fig.~\ref{fig:sed_SFR_Z}, left panel).
The right panel of Fig.~\ref{fig:sed_SFR_Z} displays the SFH of {\it Lilium}. At $z\approx 8.4$, i.e. around 120~Myr before observations, {\it Lilium} started to form stars at a progressively increasing rate. After $\approx 35$~Myr it experienced a short and intense burst of star formation ($\rm SFR\gsim 25\,\msunyr$), after which 80\% of the final stellar mass was already in place. Then, the SFR was gradually suppressed by SNe, which are the dominating feedback mechanism in our simulations. Indeed, the galaxy continued to form stars at a much lower rate, $\lsim 2\,\msunyr$, for $\approx 50$~Myr, i.e. up to $z\approx 7.45$ when it was completely quenched. Therefore, at the redshift of observation, $z = 7.3$, the average stellar age of {\it Lilium} is $\approx 90$~Myr old (see colorbar in Fig.~\ref{fig:dutycycle}).

In the same panel, we plot a SFH similar to the one derived by \cite{Looser23} using \code{BAGPIPES}, essentially a top-hat with a $\rm SFR=15~\msunyr$ but with the same $\Delta t_{\rm quench}$ and $M_\star$ as {\it Lilium}. With this SFR, the observed $M_\star$ is produced in only $\approx 30$~Myr, resulting in a much younger ($\approx 30$~Myr) stellar population. The other SED-fitting codes\footnote{The code \code{BEAGLE} \citep{chevallard_beagle} is also used in \cite{Looser23}. However, since it does not provide a reconstruction of the SFH or the average stellar age, a direct comparison with this model is not possible.} used by \cite{Looser23}, \code{prospector} \citep{prospector21} and \code{PXFF} \citep{cappellari17ppxf}, also draw similar conclusions, resulting in average stellar ages always lower than $\lesssim50$~Myr.
As a consequence, the corresponding experimentally-derived SED reaches fluxes about 10$\times$ higher than {\it Lilium} (Fig.~\ref{fig:sed_SFR_Z}, left) and matches JADES-GS-z7-01-QU photometry if a metallicity of $0.01\, \zsun$ is further assumed.

This result shows that the discrepancy between the SEDs of {\it Lilium} and JADES-GS-z7-01-QU is largely due to differences in their SFHs, and to a lesser extent, metallicity. The key difference is the fact the SFR decrease in \textit{Lilium} after the SF peak is too prolonged, resulting in a  high fraction of old stars at the time of the observation. This suggests the need for an abrupt quenching right after the SF peak.

To test this hypothesis, we artificially impose that SF in \textit{Lilium} was abruptly halted $\sim 5$~Myr after the peak, and we renormalize it to get the same $M_\star$ value.
We then assume to observe the galaxy after $\Delta t_{\rm quench}=10-20$~Myr from the halt, and thus shift the SFHs to match the redshift of the observations, i.e. $z=7.3$. As a consequence of the overall galaxy lifetime time being shorter, the duty cycle of \textit{Lilium} lowers to $\approx 0.7$.
We find that such modefied \textit{Lilium} SED now perfectly matches the observed one even by assuming the simulated stellar metallicity, $Z_\star \approx \zsun$.

\section{Discussion}\label{sec:summary}

In spite of the fact that \textit{Lilium} is very similar to JADES-GS-z7-01-QU, the simulated SED does not match the observed one. We have shown that this is due to the fact that star formation in JADES-GS-z7-01-QU is quenched on a much shorter time scale than in \textit{Lilium}.

If the quenching has to be produced by SNe, there is an intrinsic timescale on which such feedback acts, which is given by the time over which SNe associated with a given burst explode. This value is $\simgt 30$ Myr for a Kroupa IMF (see, e.g. Fig.~2 in \citealt{Pallottini17}). The quenching cannot therefore be significantly shorter than this minimum value, and this is exactly what we see in the SFH of \textit{Lilium} and the other quenched low-mass galaxies (Q1, Q2, Q3, Q4).
Basically, the SFR decline under the action of SN feedback is very gradual, resulting in a too large a number of stars that are old by the time when the galaxy is observed. This is a general problem of SF quenching in low-mass galaxies, which inherently depends on the delay of SN explosions associated to the deaths of progenitors of different mass \citep[e.g.][]{Rosdahl17}.

We speculate that the decline could be more abrupt if the mechanical energy is instead provided by a hidden AGN and/or radiation pressure from young massive stars \citep[e.g.][]{Carniani16,Ferrara23,Ziparo23}.
Indeed we may in general estimate the quenching timescale as $\Delta t_{\rm quench} = t_{\rm delay} + t_{\rm ej}$, where $t_{\rm delay}$ is the delay time associated with the onset of the physical process causing the quenching and $t_{\rm ej}$ is the ejection time required for the outflow to drive the gas out of the galaxy.
While for SNe we have $t_{\rm delay} \gtrsim \rm 30~Myr$, for radiation-driven feedback there is no such delay and we can evaluate the overall quenching timescale simply as $\Delta t_{\rm quench} = t_{\rm ej}$.
Assuming a typical outflow velocity of $v>\rm 200~km/s$ and requiring to expel the gas at a distance of $d\approx500\rm~pc$ to quench star formation (given that the effective radius of the galaxy is $\approx100\rm~pc$), we obtain:
$\Delta~t_{\rm quench}  = d/v \rm \lesssim \frac{500~pc}{200 km/s} = 2.4~Myr$.
The gas in the galaxy is expected to be set in motion and removed in such a short time as soon as the galaxy exceeds the Eddington limit. Specifically this can happen through:
i) stellar radiation, when the specific SFR exceeds a threshold value ($>13 \rm~Gyr^{-1}$, see \citealt{Fiore23})
ii) AGN feedback from massive black holes that may be a viable solution for high-z galaxies according to recent results \citep[e.g.][]{Maiolino23_gnz11}.
We therefore conclude that the SFR decline might be a powerful diagnostic of different feedback types.

Another puzzle posed by the JADES-GS-z7-01-QU observation is its long latency before the first episode of star formation. This source is observed at $z=7.3$ and, according to the SED interpretation, has formed all its stars in a short ($\approx 30\, \rm Myr$) burst. This sets the start of its star formation activity at $z \approx 7.8$. From the observed stellar mass $M_\star=10^{8.7} \msun$, and conservatively assuming that all its baryons were turned into stars, we can set a lower limit to the host halo, $M\simgt (\Omega_m/\Omega_b) M_\star = 3.2\times 10^9 \msun$. Assuming a standard growth history, this halo should have crossed the critical mass $M\approx 10^8 \msun$, marking the separation between minihalos, in which the SF is easily suppressed, and the star-forming Ly$\alpha$-cooling halos at $z\simgt 11.7$. Thus, in the redshift interval $7.8 < z < 11.7$ (280 Myr), although conditions were in principle favourable to star formation in terms of gas cooling, the galaxy did not form a significant amount of stars.
A possibility might be the following: if a galaxy spawns a massive stellar cluster ($M_\star \simeq 10^5 \msun$) as its first star formation event, its radiation can photo-dissociate molecular hydrogen and prevent further star formation for up to $\sim 300$~Myr \citep[see Alyssum,][]{Pallottini2022}.

As JWST will build a sizeable sample of quiescent high-$z$ galaxies, we will be able to solve these puzzles and understand the nature of the feedback-regulated evolution of low-mass systems.

\begin{acknowledgments}
This project received funding from the ERC Starting Grant NEFERTITI H2020/804240 (PI: Salvadori). AF, AP and SC acknowledge the ERC Advanced Grant INTERSTELLAR H2020/740120 (PI: Ferrara). We acknowledge the CINECA award under the ISCRA initiative, for the availability of high performance computing resources and support from Class B project SERRA HP10BPUZ8F (PI: Pallottini) and the computational resources of the Center for High Performance Computing (CHPC) at SNS.
\end{acknowledgments}

\bibliography{refer,codes}
\bibliographystyle{aasjournal}
\end{document}